\documentclass{article}
\usepackage{graphicx}
\begin{document}
\baselineskip 25pt

\title{Crystal structures, magnetic and superconducting properties of the
RuSr$_2$NdCu$_2$O$_x$ and RuSr$_2$GdCu$_2$O$_y $ compounds}

\maketitle

\author{A. Vecchione,$^a $\footnote{\noindent Antonio Vecchione, Unit\`a I.N.F.M. di Salerno, Dipartimento di
Fisica \lq\lq E.R. Caianiello\rq\rq, Universit\`a di Salerno, Via
Salvador Allende, I-84081 Baronissi (Salerno), Italy. Fax: +39 089
953804, e-mail: vecchione@sa.infn.it}
M. Gombos,$^a$ S. Pace,$^a$ C. Tedesco,$^b$ and D.Zola.$^a$}

\vskip 15pt
\centerline{$^a$ Unit\`a I.N.F.M. di Salerno,
        Dipartimento di Fisica \lq\lq E.R. Caianiello\rq\rq}
\centerline{Universit\`a di Salerno, Via S. Allende, I-84081 Baronissi (Salerno), Italy}

\centerline{$^b$ Dipartimento di Chimica, Universit\`{a} di Salerno,}
\centerline{ via S. Allende, I-84081 Baronissi, Italy}

%\thanks{ \noindent Antonio Vecchione, Unit\`a I.N.F.M. di Salerno, Dipartimento di
%Fisica \lq\lq E.R. Caianiello\rq\rq, Universit\`a di Salerno, Via
%Salvador Allende, I-84081 Baronissi (Salerno), Italy. Fax: +39 089
%953804, e-mail: vecchione@sa.infn.it}

%\maketitle \draft

\begin{abstract}

We report the magnetization and the susceptibility measurements of
the RuSr$_2$GdCu$_2$O$_y$ and RuSr$_2$NdCu$_2$O$_x$ perovskite
materials. We find that RuSr$_2$GdCu$_2$O$_y$ compound exhibits a
magnetic transition at $T_n$=135 K followed by a superconducting
one with an onset $T_c$=35 K. Samples of RuSr$_2$NdCu$_2$O$_x$
material have shown neither superconductivity nor magnetic
transition. XRD measurements show that in RuSr$_2$NdCu$_2$O$_x$
the Nd ions tend to substitute to Sr ions in very high
percentages. The relevance of this phenomenon on the absence of
superconductivity in RuSr$_2$NdCu$_2$O$_x$ is discussed.


\end{abstract}

\newpage

The rutheno-cuprate RuSr$_2$GdCu$_2$O$_y$ (Gd-1212) system is a
subject of great interest because exhibits coexistence of bulk
superconductivity and magnetic ordering \cite{tallon1}. The
RuSr$_2$GdCu$_2$O$_y$ \hfill tetragonal \hfill structure \hfill is
an \hfill analog \hfill of \break
Y$_1$Ba$_2$Cu$_3$O$_7$ with replacement of Ru ions into Cu(1)
sites. This results in a perovskite consisting of different
sequences of alternating CuO$_2$ bilayers and RuO$_2$ monolayer.
In this hybrid rutheno-cuprate system both the Cu-O and Ru-O
planes form very similar square-planar arrays, and the coexistence
of superconductivity and long range magnetic order is intriguing
\cite{tallon2}. However, in spite of extensive investigation a
consistent picture of the magnetic structure is still lacking
\cite{tallon3},\cite{tallon4}.

\noindent In the present investigation, we show the effects on the
structural and superconducting properties  of Gd-1212
rutheno-cuprate compound by substituting Gd rare earth with Nd.

\noindent Policrystalline samples of RuSr$_2$NdCu$_2$O$_x$ and
RuSr$_2$GdCu$_2$O$_y$ were prepared by conventional solid state
reaction by mixing stoichiometric amounts of highly pure RuO$_2$,
Nd$_2$O$_3$, Gd$_2$O$_3$, CuO oxides and strontium carbonate
(SrCO$_3$). After calcination in air at 960$^\circ$C the powders
were ground, milled and annealed in argon flow at 1020$^\circ$C
for 10 hours. At the final stage the powders were oxygenated with
a procedure made by several cycles of heating at temperatures
ranging from 1050$^\circ$C to 1080 $^\circ$C. The phases produced
after each step were monitored by X-ray diffraction.

%\begin{figure}[h]
%\includegraphics[width=17.5pc]{c:\varie\lavori\eucas2001\cell.eps}
%\caption{\label{fig1}X-ray diffraction powder patterns of the
%RuSr$_2$NdCu$_2$O$_x$ (up) and RuSr$_2$GdCu$_2$O$_y$ (bottom).
%Peaks marked by stars are due to phase impurities.}
%X-ray diffraction powder patterns of the
%RuSr$_2$NdCu$_2$O$_x$ (up) and RuSr$_2$GdCu$_2$O$_y$ (bottom).
%Peaks marked by stars are due to phase impurities
%\end{figure}

%\begin{figure}[h]
%\includegraphics[width=17.5pc]{c:\varie\lavori\eucas2001\cell.eps}
%\caption{\label{fig2} DC susceptibility measurement of the
%RuSr$_2$NdCu$_2$O$_x$ compound. Inset: 1/$\chi$ versus $T$ showing
%Curie paramagnetic behaviour.}
%\end{figure}

\noindent X-ray diffraction patterns were recorded by means of
Philips PW-1700 diffractometer using Ni-filtered Cu K$\alpha$
radiation. Magnetization and susceptibility measurements were
performed by means of a Oxford Maglab Vibrating Sample
Magnetometer (VSM). The samples vibrate in a region where the
field homogeneity is equal to ${10^{-6}}$. Structural and
magnetization measurements were carried out after sieving the
materials and collecting the fraction with size smaller than 10
$\mu$m. The correct stoichiometry of both RuSr$_2$NdCu$_2$O$_x$
and RuSr$_2$GdCu$_2$O$_y$ compounds and the absence of
contamination from both alumina crucibles and other spurious
elements were checked by EDS analysis.

\noindent X-ray \hfill diffraction \hfill patterns \hfill of \hfill oxygen \hfill annealed
\hfill RuSr$_2$NdCu$_2$O$_x$\hfill and \break RuSr$_2$GdCu$_2$O$_y$ samples are shown
in Fig.1. Reflections marked by stars are due to small amounts of
extra-phase. X-ray diffraction patterns of RuSr$_2$NdCu$_2$O$_x$
has been successfully indexed assuming a cubic cell with $a$=3.907
\AA. This implies that a disordered cubic perovskite structure can
be assumed: Nd and Sr cations occupy the same site and the same
applies to Cu and Ru. As already reported \cite{tallon2},
RuSr$_2$GdCu$_2$O$_y$ exhibits a tetragonal cell  ($a$=3.838 \AA,
$c$=11.573 \AA) with alternating CuO$_2$ and RuO$_2$ planes. The
proposed structural model for Nd-1212 compound in space group
$Pm{\overline3}m$ is similar to that of RuSr$_2$GdCu$_2$O$_y$ with
mixed Cu/Ru and Nd/Sr cation sites and oxygen coordinates fixed by
the symmetry of the space group.

\noindent The possible effects of the cations intermixing on the
magnetic and superconducting properties were analyzed by
susceptibility and magnetization measurements. We measured \hfill a DC
\hfill susceptibility, $\chi$, for \hfill RuSr$_2$NdCu$_2$O$_x$ \hfill and \break
RuSr$_2$GdCu$_2$O$_y$ samples in the temperature range from 5 K to
270 K. In Fig.2 we show the zero-field-cooled DC susceptibility,
$\chi(T)$, measured in RuSr$_2$NdCu$_2$O$_x$ powders in an
external field of 10 G. In the temperature range investigated
neither ferromagnetic nor superconducting transition occurs. The
RuSr$_2$NdCu$_2$O$_x$ is paramagnetic and follows a behaviour well
described in term of Curie law for $T<$ 150 K (see inset of
Fig.2).  We also carried out measurements of $\chi (T)$, shown in
the inset of the Fig.3, on a RuSr$_2$GdCu$_2$O$_y$ sample. The
main features observed are a large rise for $T <$ 150 K due to the
magnetic transition and a large drop for $T <$ 35 K ascribed to
the onset superconducting transition. The signal revealed in
superconducting region is not diamagnetic and no Meissner phase is
observed down to 5 K as already reported in literature
\cite{tallon1,chu}. We do not observe a peak in DC susceptibility.
A reason could be in our experimental set-up. In fact the
superconducting magnet of the VSM has a persistent field measured
by Hall probe which has 1 G as sensibility. To perform zero field
cooling measurement we cancel this persistent field applying an
opposite magnetic field. Probably the residual field lower than 1
G is enough to orientate the magnetic domain in our sample during
the initial cooling. However, the peak in zero field cooled
susceptibility was not observed by Felner  et al. in not
superconducting RuSr$_2$EuCu$_2$O$_8$ and by Williams and Kramer
for field higher than 2.5 kG \cite{felner, williams}.

\noindent In order to investigate the magnetic behaviour in the
RuSr$_2$GdCu$_2$O$_y$ sample, we performed measurements of the
magnetization $M$ in external magnetic field $H$ up to 5 T at
different temperatures. As can be observed from Fig. 3, an
irreversible behaviour is revealed in $M(H)$ loop. This finding
has been ascribed \cite{tallon1} to the presence of a
ferromagnetic phase in these samples.

%\begin{figure}[h]
%\includegraphics[width=17.5pc]{c:\varie\lavori\eucas2001\cell.eps}
%\caption{\label{fig3} Magnetization measurements of
%RuSr$_2$GdCu$_2$O$_y$ taken at three different temperatures.
%Inset: DC susceptibility versus $T$ measurement.}
%\end{figure}

\noindent Our data point out that the substitution of the
neodymium in Gd-1212 compounds involves a change in the
superconducting and magnetic properties. The observed magnetic
transition in RuSr$_2$GdCu$_2$O$_y$, attributed to the ordering in
the ruthenium planes, does not take place in RuSr$_2$NdCu$_2$O$_x$
. At the same time the superconducting transition, occurring in
copper planes of Gd-based compound, is also suppressed. The
presence of structural disorder in Nd-based compound play a role
preventing a long range extension of both Cu-O and Ru-O planes
through out the structure. This implies the suppression of the
superconducting and (ferro)magnetic order parameters.

\noindent In conclusion, we have prepared RuSr$_2$NdCu$_2$O$_x$
and RuSr$_2$GdCu$_2$O$_y$ materials whose structure and transport
properties have been described by  XRD analysis and magnetization
and susceptibility measurements. A description of the x-ray
patterns has been achieved assuming that the  Nd-based 1212 phase
is cubic and  not isostructural to the Gd-1212 phase. In
particular, we attribute the absence of superconductivity and
magnetic ordering in RuSr$_2$NdCu$_2$O$_x$ compound to the partial
intermixing between Ru and Cu and between Nd and Sr into the
crystalline structure.
\\ We acknowledge Dr. Canio Noce for helpful discussions.

\newpage

\centerline{\bf Figure Captions}

\vskip 20pt
{\bf Figure 1.} X-ray diffraction powder patterns of the
RuSr$_2$NdCu$_2$O$_x$ (up) and RuSr$_2$GdCu$_2$O$_y$ (bottom).
Peaks marked by stars are due to phase impurities.

\vskip 20pt

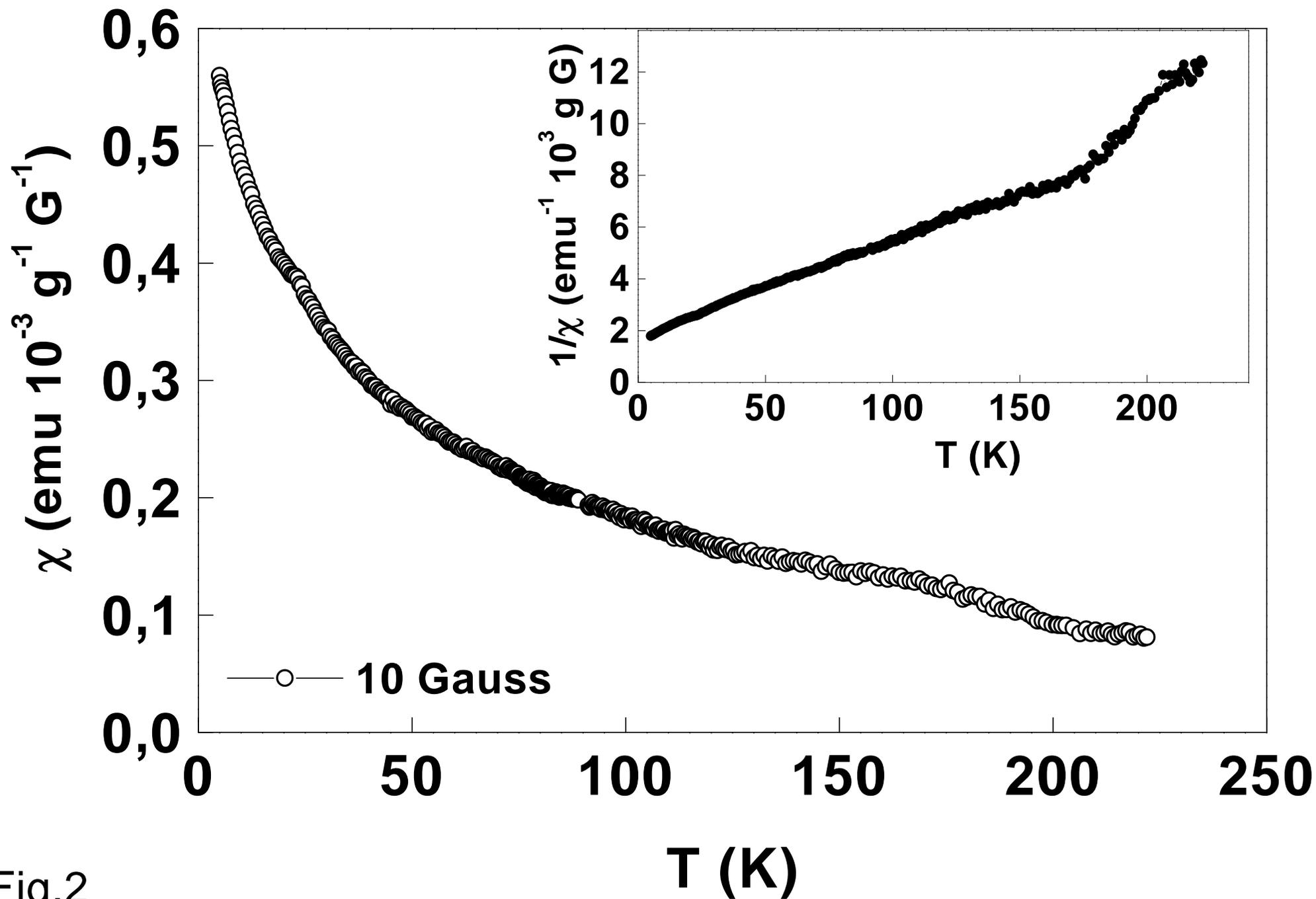
{\bf Figure 2.} DC susceptibility measurement of the
RuSr$_2$NdCu$_2$O$_x$ compound. Inset: 1/$\chi$ versus $T$ showing
Curie paramagnetic behaviour.

\vskip 20pt

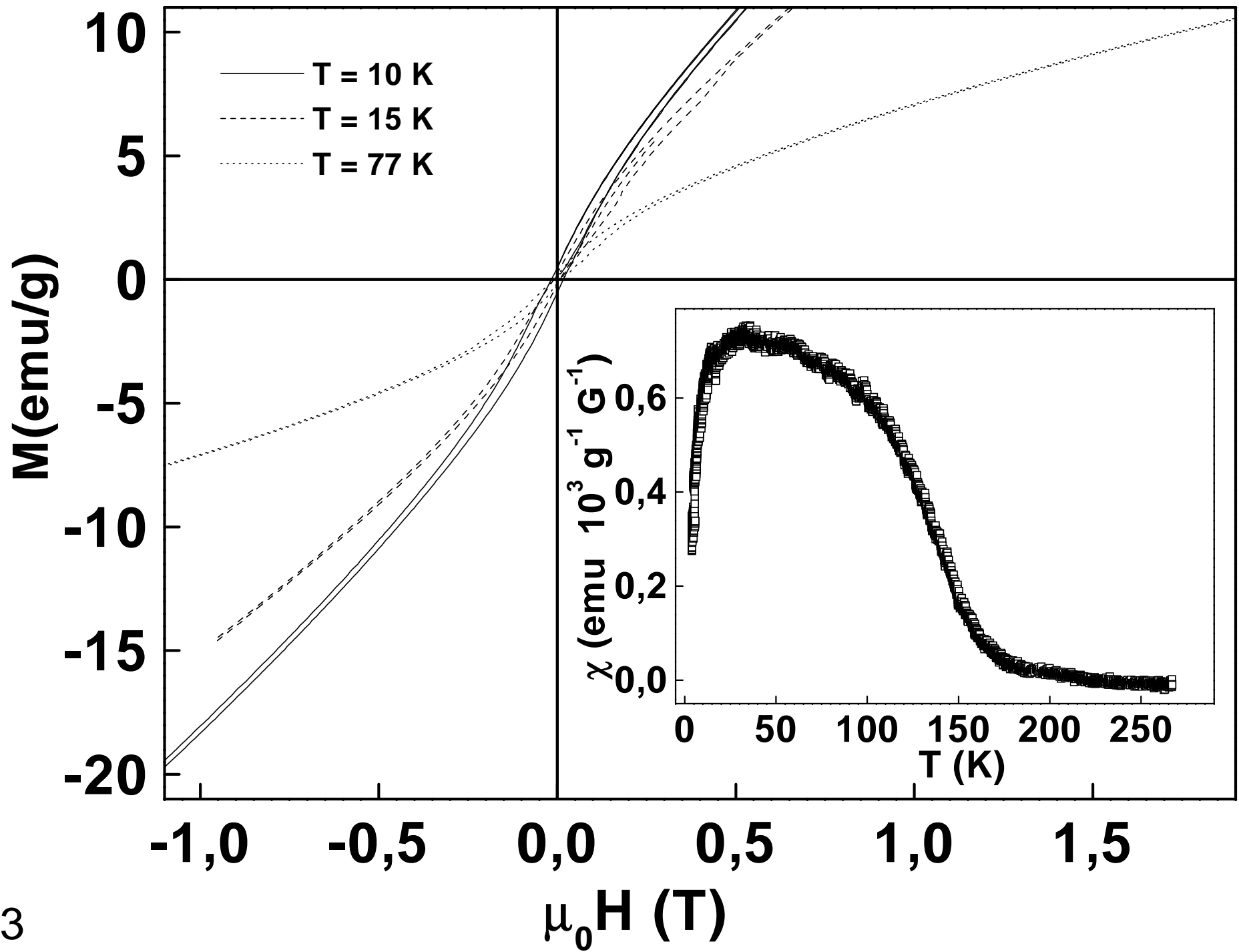
{\bf Figure 3.} Magnetization measurements of RuSr$_2$GdCu$_2$O$_y$
taken at three different temperatures. Inset: DC susceptibility
tversus $T$ measurement.


\begin{thebibliography}{9}

\bibitem{tallon1} C. Bernhard {\it et al.}, Phys. Rev. B {\bf 59}, 14099 (1999).
\bibitem{tallon2} A. C. McLaughlin {\it et al.}, Phys. Rev. B {\bf 60}, 7512 (1999).
\bibitem{tallon3} A. Fainstein {\it et al.}, Phys. Rev. B {\bf 60}, 12597 (1999);
A. C. McLaughlin and J. P. Attfield, Phys. Rev. B {\bf 60}, 14605
(1999).
\bibitem{tallon4} J.W. Lynn {\it et al.},
Phys Rev B {\bf  61}, 14 964 (2000).
\bibitem{chu}C.W. Chu et al. Physica C {\bf335}, 231 (2000)
\bibitem{felner}I. Felner, U. Asaf, Y. Levi, and O. Millo, Phys. Rev. B {\bf 55}, 3374
(1997); E. B. Sonin and I. Felner, Phys. Rev. B {\bf 57}, 14000
(1998); I. Felner, U. Asaf, S. Reich, and Y. Tsabba, Physica C
{\bf 311}, 163 (1999); I. Felner, U. Asaf, Y. Levi, and O. Millo,
Physica C {\bf 334}, 141 (2000).
\bibitem{williams}G. V. M. Williams and S. Kr$\ddot{a}$mer, Phys. Rev. B
{\bf 62}, 4132
(2000).


\end{thebibliography}
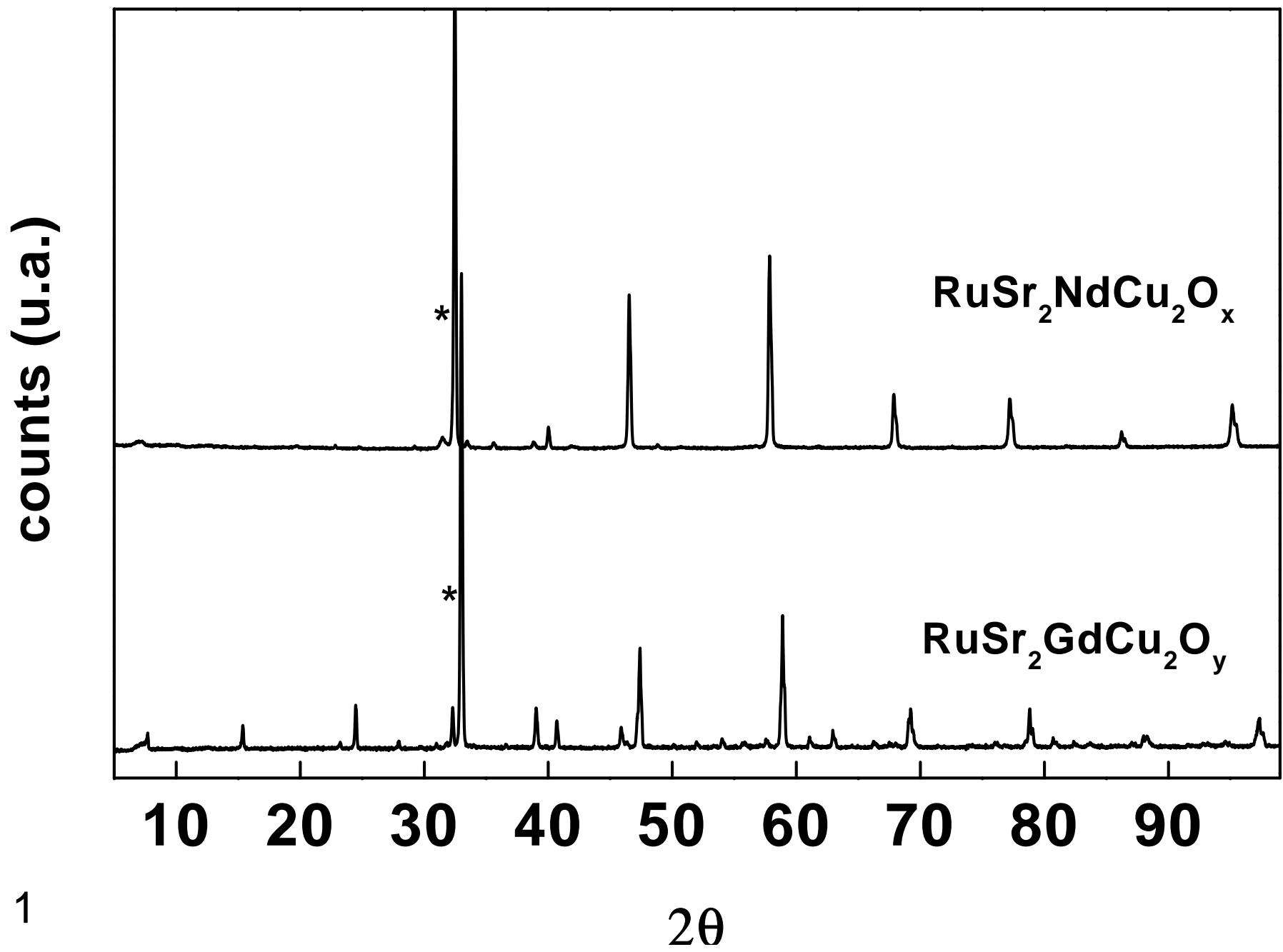
\end{document}